\documentclass[graybox]{svmult}

\usepackage{type1cm}        
\usepackage{makeidx}         
\usepackage{graphicx}        
\usepackage{multicol}        
\usepackage[bottom]{footmisc}

\usepackage{newtxtext}       %
\usepackage[varvw]{newtxmath}       

\usepackage[numbers,sort&compress]{natbib}

\newcommand{\hl}[1]{{\color{black}#1}}

\makeindex             

\begin{document}

\title*{A glitch in gravity: cosmic Lorentz-violation from fiery Big Bang to glacial heat death}
\titlerunning{A glitch in gravity} 
\authorrunning{Wen, Hergt, Afshordi\& Scott} 

\author{Robin Y. Wen \and \\ Lukas T. Hergt \and \\ Niayesh Afshordi  \and \\ Douglas Scott}
\institute{Robin Y. Wen \at California Institute of Technology, Pasadena, CA 91125, USA \\ \email{ywen@caltech.edu}
\and Lukas T. Hergt \at Department of Physics \& Astronomy, University of British Columbia, Vancouver, BC V6T 1Z1, Canada \\ \email{lthergt@phas.ubc.ca}
\and Niayesh Afshordi \at Waterloo Centre for Astrophysics, University of Waterloo, Waterloo, ON, N2L 3G1, Canada \\ Department of Physics \& Astronomy, University of Waterloo, Waterloo, ON N2L 3G1, Canada \\ Perimeter Institute For Theoretical Physics, Waterloo, ON N2L 2Y5, Canada \\ \email{nafshordi@pitp.ca}
\and Douglas Scott \at Department of Physics \& Astronomy, University of British Columbia, Vancouver, BC V6T 1Z1, Canada \\ \email{dscott@phas.ubc.ca}
}
%
%

\maketitle
{\it To be published in Astrophysics and Space Science Proceedings, titled ”The Relativistic Universe: From Classical to Quantum, Proceedings of the International Symposium on Recent Developments in Relativistic Astrophysics”, Gangtok, December 11-13, 2023: to felicitate Prof. Banibrata Mukhopadhyay on his 50th Birth Anniversary”, Editors: S Ghosh \& A R Rao, Springer Nature; Reviewing and extending \cite{Wen:2023wes}}
\\
\\
\\

\abstract{One regime where we might see departures from general relativity is at the largest accessible scales, with a natural choice in cosmology being the cosmological horizon (or Hubble) scale. We investigate a single-parameter extension to the standard cosmological model with a different strength of gravity above and below this scale---a ``cosmic glitch'' in gravity. Cosmic microwave background observations, and Baryonic Acoustic Oscillations (including the recent DESI Y1)  favour weaker superhorizon gravity, at nearly a percent (or 2$\sigma$ level), easing both the Hubble and clustering tensions with other cosmological data. This compounds evidence for an even stronger glitch during Big Bang nucleosynthesis (from helium abundance observations), suggesting that symmetries of general relativity are maximally violated at the Big Bang, but gradually recovered as we approach the present-day cosmological de Sitter scale, associated with the observed dark energy.}


\section{Introduction}

During the last century, the general theory of relativity (GR) has become a fundamental pillar of modern physics, successfully passing every empirical examination across a wide range of scales and regimes~\cite{will2020einstein}. However, several reasons, most importantly the inconsistency of the principles of general relativity with those of quantum theory, have motivated extensive studies on deviations or modifications beyond GR. Additionally, the presence of a preferred cosmological reference frame, in which the cosmic microwave background~(CMB) anisotropies possess no dipole, could point towards alternative gravity theories that can allow a genuinely unique cosmological frame. The existence of such a special frame could imply that minimal Lorentz-violating deviations from GR would only appear on cosmological scales~\cite{07CuscutonCosmo,mukohyama2019minimally}. 

Instead of requiring a new fundamental length scale, an alternative implementation of a minimal Lorentz violation at cosmological distances is to introduce a ``glitch'' between Newton's constant of gravitation~$G_\mathrm{N}$ that determines the sub-horizon dynamics and the gravitational constant~$G_\mathrm{cosmo}$ that affects cosmology on super-horizon scales.  More specifically, we can write the Friedmann equation as
\begin{equation}
    H^2= \frac{8\pi G_\mathrm{cosmo}}{3}\rho_\mathrm{tot}=\frac{8\pi G_\mathrm{N}}{3}\left[\rho_\mathrm{tot}+\left(\frac{G_\mathrm{cosmo}}{G_\mathrm{N}}-1\right)\rho_\mathrm{tot}\right]\,,
    \label{eq:CGG-1}
\end{equation}
where $G_\mathrm{cosmo}$ connects the total energy density~$\rho_\mathrm{tot}$ in the Universe to the Hubble expansion rate~$H$. Such a cosmic glitch can be realised through alternative gravity theories such as cuscuton and Ho\v{r}ava--Lifshitz proposals for Lorentz-violating gravity~\cite{hovrava2009quantum,09CuscutonHovrava} or the Einstein-Aether theory~\cite{jacobson2010extended}, with the ratio $G_\mathrm{N}/G_\mathrm{cosmo}$ depending on the theories' respective parameters ($G_\mathrm{N}/G_\mathrm{cosmo}=1$ in GR without a ``glitch''). It can be rigorously shown that all of these theories in appropriate limits are indistinguishable from general relativity in asymptotically flat spacetimes~\cite{Loll:2014xja}, so they can only be examined on scales that are close to or larger than the Hubble scale~\cite{Robbers:2007ca}.


\section{The glitch model}

Phenomenologically, this cosmic glitch in gravity (hereafter referred to as CGG) can be reformulated as an additional dark energy component on top of the usual cosmological constant~$\Lambda$ and cold-dark-matter fluids within the concordance model of cosmology (commonly known as $\Lambda$CDM). Analogously to how the cosmological constant~$\Lambda$ is typically reinterpreted as a vacuum energy density~$\rho_\Lambda=\Lambda/8\pi G_\mathrm{N}$ in the $\Lambda$CDM model under GR, we assume that the second term in the brackets of \ref{eq:CGG-1} corresponds to some glitch energy density~$\rho_\mathrm{g}$ that contributes to the overall critical density $\rho_\mathrm{crit}\equiv3H^2/8\pi G_\mathrm{N}$ of GR. From {\ref{eq:CGG-1}, the total density can be expressed in terms of the critical density,
\begin{align}
    \rho_{\rm tot}=\frac{G_\mathrm{N}}{G_\mathrm{cosmo}}\rho_{\rm crit}
    \label{eq:crit-tot}\,.
\end{align}
Substituting the above \ref{eq:crit-tot} into the second term of \ref{eq:CGG-1}, we obtain}
\begin{equation}
    H^2= \frac{8\pi G_\mathrm{N}}{3}\left[\rho_\mathrm{tot}+\left(1-\frac{G_\mathrm{N}}{G_\mathrm{cosmo}}\right)\rho_\mathrm{crit}\right].
    \label{eq:CGG-2}
\end{equation}
We can reinterpret the CGG (the second term of the above eq.~\ref{eq:CGG-2}) as an additional dark energy component with a constant parameter proportional to the critical density:
\begin{equation}
    \Omega_\mathrm{g} 
    \equiv \frac{\rho_\mathrm{g}}{\rho_\mathrm{crit}} 
    = \frac{\rho_\mathrm{DE} -\rho_\Lambda}{\rho_\mathrm{crit}}
    = 1 -\frac{G_\mathrm{N}}{G_\mathrm{cosmo}}.
    \label{eq:G_cosmo}
\end{equation}
The total \emph{effective} dark energy component now has the energy density constituted by the combination of the constant dark energy density (the cosmological constant) and the energy density of the glitch:
\begin{equation}
    \rho_\mathrm{DE}= \rho_{\Lambda}+\Omega_\mathrm{g}\rho_\mathrm{crit}.
    \label{eq:density-CGG}
\end{equation}
Given that $\rho_\mathrm{crit}$ also contains $\rho_\mathrm{DE}$, we can rearrange \ref{eq:density-CGG} to isolate $\rho_\mathrm{DE}$:
\begin{equation}
    \rho_\mathrm{DE}= \frac{\Omega_\mathrm{g}\rho_\mathrm{nonDE}+\rho_{\Lambda}}{1-\Omega_\mathrm{g}},
    \label{eq:density-CGG-2}
\end{equation}
where $\rho_\mathrm{nonDE}$ includes the densities of all the components other than the dark energy. As $\rho_{\Lambda}$ becomes dominant over $\rho_\mathrm{nonDE}$ in the late Universe, the behavior of this modified dark energy model closely resembles that of $\Lambda$CDM during late times. However, the dynamics of $\rho_\mathrm{DE}$ undergo considerable changes in the early Universe due to its proportionality with the critical density; the modified dark energy component tracks the behavior of radiation and matter densities, respectively, when the radiation and matter dominate over the total energy budget of the Universe.

\begin{figure}[tbp]
    \centering
    \includegraphics[width=\textwidth]{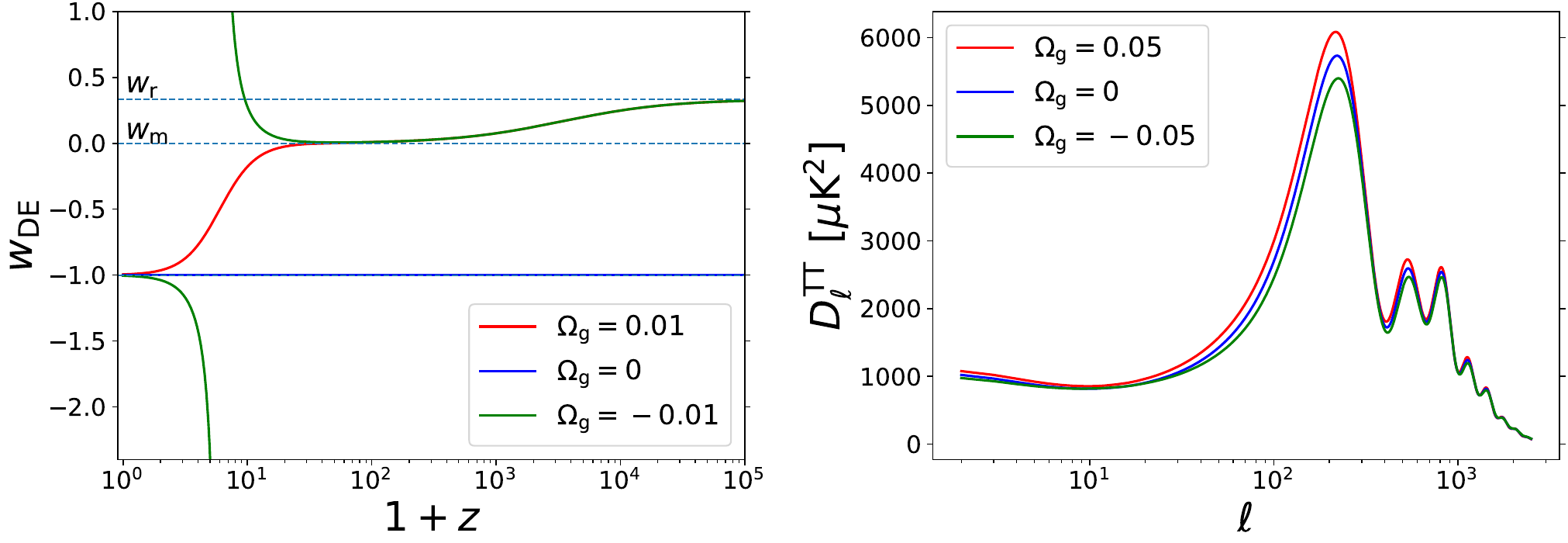}
    \caption{The cosmic glitch in gravity (CGG) model. The left panel shows the EoS parameter for the dark energy component $w_\mathrm{DE}(z)$ for different values of the glitch parameter~$\Omega_\mathrm{g}$ (with all other cosmological parameters fixed at the best-fit values for the Planck18 data under the $\Lambda$CDM model). The blue dashed lines plot the EoS parameters for radiation ($w_\mathrm{r}=\frac{1}{3}$) and matter ($w_\mathrm{m}=0$). The right panel shows the CMB temperature power spectrum $D_{\ell}^{TT}\equiv(\ell(\ell+1)/2\pi)C_{\ell}^{TT}$ for different glitch parameters.}
    \label{fig:w-omega}
\end{figure}


\begin{figure}[tbp]
    \centering
    \includegraphics[width=0.53\textwidth]{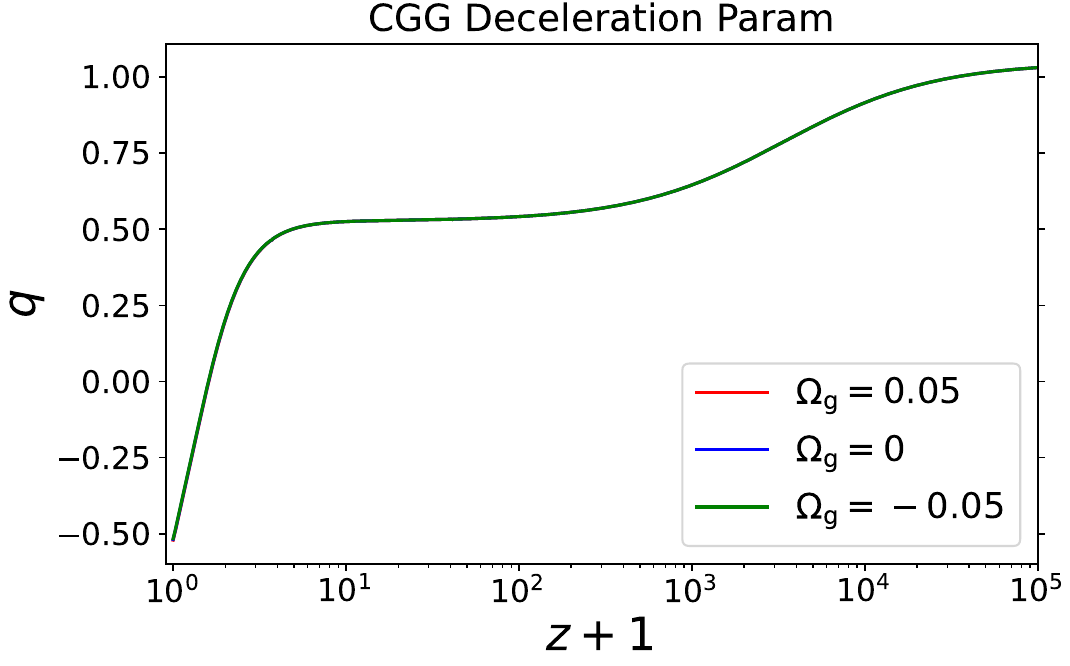}
    \caption{{The deceleration parameter $q$ of the CGG model for different $\Omega_{\rm g}$ values.}}
    \label{fig:deceleration}
\end{figure}

Using~\ref{eq:density-CGG-2}, we can compute the equation of state (EoS) parameter $w_\mathrm{DE}$ (defined as the ratio between pressure and energy density~$w\equiv p/\rho$) for the effective dark energy component and confirm that the CGG model tracks the dominant component contributing to the Universe's density. As plotted in the left panel of~\ref{fig:w-omega}, $w_\mathrm{DE}$ evolves from $\frac{1}{3}$ (EoS for radiation) to $0$ (EoS for matter) and then to around $-1$ (EoS for the constant dark energy $\Lambda$). The EoS parameter~$w_\mathrm{DE}$ changes smoothly when $\Omega_\mathrm{g}$ is positive. However, for negative~$\Omega_\mathrm{g}$ values, $w_\mathrm{DE}$ becomes divergent at the positive-to-negative transition of $\rho_\mathrm{DE}$. A negative energy density~$\rho_\mathrm{DE}$ typically leads to ghost instabilities in dynamical dark energy models \cite{11QCD_ghost,15Fulvio_ghost}, but here we have formulated the model by taking a negative glitch value~$\Omega_\mathrm{g}$ to be a phenomenological approximation to modified gravity theories. This means that such Lorentz-violating theories do not have new dynamical degrees of freedom that could suffer from ghost instabilities~\cite{07CuscutonCosmo,mukohyama2019minimally}. {We also note that the divergence of $\omega_{\rm DE}$ in the case of negative $\Omega_{\rm g}$ does not lead to any divergent behavior in the deceleration parameter ($q\equiv -\ddot{a}a/\dot{a}^2=\frac{1}{2}\sum_{i}\Omega_i(1+3\omega_i)$, where the sum occurs over all the cosmological components), as shown in \ref{fig:deceleration}. In fact, the deceleration parameter remains the same across time, despite varying $\Omega_{\rm g}$ values, indicating that the CGG model does not change the background dynamics at or beyond second order.}

We can confront our CGG model with cosmological data to see how well it fits. To compute the CMB anisotropy power spectra under the CGG model, we modify the cosmological Boltzmann code \texttt{CAMB}~\cite{CAMB}. We treat the CGG component as a perfect fluid at the linear perturbation level. For negative $\Omega_\mathrm{g}$, the model crosses the so-called ``phantom~divide'', with $w<-1$ at late times. To ensure gravitational stability at such a crossing, a double-field description,\footnote{A single scalar field with minimal coupling will be insufficient for stability.} known as the parameterized post-Friedmann (PPF) framework, has been used to solve the DE perturbation equations~\cite{08HuPPFDE,08FangPPFDE}, allowing for a generic dark energy sector with an arbitrary EoS parameter $w(z)$ as a function of redshift. We use the PPF framework for numerical implementation when $\Omega_\mathrm{g}$ takes negative values, {and we assume the sound speed of the effective dark energy fluid in the rest frame to be approximately the speed of light, that is $c_\mathrm{s}^2=1$ \cite{08FangPPFDE}}. The right panel of~\ref{fig:w-omega} illustrates the effects of $\Omega_\mathrm{g}$ on the CMB temperature power spectrum~$D_{\ell}^{TT}$: a positive $\Omega_\mathrm{g}$ enhances the integrated Sachs-Wolfe (ISW) effect at large scales (corresponding to low angular multipoles $\ell$) and slightly suppresses the small-scale power in $D_{\ell}^{TT}$, while a negative $\Omega_\mathrm{g}$ exhibits the exactly opposite behavior.

\begin{table}[tbp]
\setlength{\tabcolsep}{10pt}
\begin{center}
\begin{tabular}{l c c c}
\hline
\hline
\noalign{\vskip 1pt}
Parameter & Planck18 [$\Lambda$CDM] & Planck18 [CGG] \\
\hline
\noalign{\vskip 2pt}
$\Omega_\mathrm{b} h^2$ & $0.02237\pm0.00014$ & $0.02248\pm0.00016$ \\
$\Omega_\mathrm{c} h^2$ & $0.1200\pm0.0012$ & $0.1168\pm0.0020$ \\
$H_0$ & $67.36\pm0.54$ & $68.58\pm0.86$ \\
$\tau_\mathrm{reio}$ & $0.0542^{+0.0071}_{-0.0082}$ & $0.0499^{+0.0079}_{-0.0071}$\\
$\ln(10^{10} A_\mathrm{s})$  & $3.044\pm0.015$ & $3.033\pm 0.016$ \\
$n_\mathrm{s}$ & $0.9649\pm0.0042$ & $0.9690\pm0.0046$\\
$\Omega_\mathrm{g}$ & $0$ & $-0.0087\pm0.0046$ \\
\hline
$\Omega_\mathrm{m}$ & $0.3155\pm0.0074$ & $0.298\pm0.012$\\
$\sigma_8$ & $0.8112\pm0.0061$ & $0.831\pm0.012$ \\
$S_8\equiv\sigma_8\sqrt{\Omega_\mathrm{m}/0.3}$ & $0.832\pm0.013$ & $0.828\pm0.012$ \\
\hline
\end{tabular}
\end{center}
\caption{{Mean and $1\,\sigma$ uncertainties of the parameters in the $\Lambda$CDM and CGG models using the \textit{Planck} 2018 data. The first seven parameters are the main parameters of the models used in the nested sampling, while the last three are derived according to the seven main parameters.}}
\label{tab:Constraints}
\end{table}


\section{Observational Constraints}

In the absence of a theoretically favored value for $\Omega_\mathrm{g}$, it is best to constrain its value using the existing CMB and large-scale structure (LSS) data. Our CGG model is a single-parameter-extension to the 6-parameter $\Lambda$CDM model, and we can constrain $\Omega_\mathrm{g}$ using the \textit{Planck}~2018 $TT,TE,EE$+low$E$+lensing likelihoods (hereafter abbreviated as Planck18)~\cite{18Plancklikelihood,18Planckparameter}.\footnote{Our results were computed using the nested sampler \texttt{PolyChord}~\cite{15Handley_Polychord_stat,15Handley_polychord_cosmo} interfaced with \texttt{Cobaya}~\cite{2020Cobaya} and a modified version of \texttt{CAMB}~\cite{CAMB}.} {In \ref{tab:Constraints}, we summarize the constraints on the cosmological parameters for the $\Lambda$CDM and CGG models}\footnote{All parameter uncertainties given in this work are expressed as $\pm1\,\sigma$.}. We find that the \textit{Planck}~2018 data prefer $\Omega_\mathrm{g}$ to stay in the negative region, with $\Omega_\mathrm{g}=-0.0087\pm0.0046$. The $\Lambda$CDM model is almost $2\,\sigma$ away from the mean $\Omega_\mathrm{g}$ value measured under CGG.  
Using \ref{eq:G_cosmo}, we can recast the constraint on $\Omega_\mathrm{g}$ as a constraint on the ratio between the strengths of super-horizon and sub-horizon gravity, $G_\mathrm{cosmo}/G_\mathrm{N}=0.9914\pm0.0045$, indicating that the \textit{Planck} CMB data favor superhorizon gravity that is about a percent weaker than in GR, at the roughly $2\,\sigma$ significance level. 

{This preference for a negative glitch parameter remains at a similar significance level when we consider different constant values for the sound speed (instead of the default $c_\mathrm{s}^2=1$) of the effective dark energy components in the PPF framework. When $c_\mathrm{s}^2=0.1$, a sound speed substantially smaller the speed of light, we obtain $\Omega_\mathrm{g}=-0.0098^{+0.0055}_{-0.0050}$ under the Planck18 data, while increasing the sound speed to $c_\mathrm{s}^2=10$ yields $\Omega_\mathrm{g}=-0.0063\pm 0.0034$. We see that increasing (decreasing) the DE sound speed increases (decreases) the mean constraint on the glitch parameter. Since the $1\,\sigma$ error of the constraint decreases (increases) with a similar percentage to the mean value, the preference for the negative glitch is kept stable at around the $2\,\sigma$ level, regardless of the $c_\mathrm{s}^2$ values.}

\begin{figure}[tbp]
    \centering\includegraphics[width=1.0\textwidth]{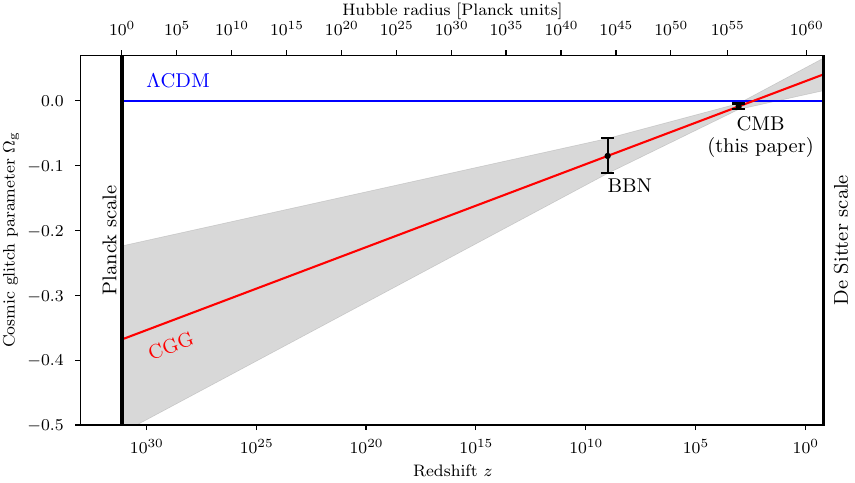}
    \caption{Current measurements of a time-varying cosmic glitch parameter $\Omega_\mathrm{g}$, as a function of Hubble radius at the measurement time. The lines and gray region show the linear extrapolation ($\pm1\,\sigma$) out to Planck and de~Sitter scales.}
    \label{fig:running}
\end{figure}

\subsection{A logarithmic running of the glitch}

The preference of the \textit{Planck} CMB data for a negative $\Omega_\mathrm{g}$ is interestingly echoed by recent measurements on Big Bang nucleosynthesis~(BBN). The EMPRESS collaboration finds an apparent discrepancy between the standard model of BBN and the measurement of $^4$He abundance in 10 extremely metal-poor galaxies \cite{matsumoto2022empress}. The BBN predictions can be reconciled with the EMPRESS data if $\Omega_\mathrm{g} = -0.085\pm0.027$ during nucleosynthesis~\cite{22EMPRESSVIIICuscuton}. However, this value needed to explain the abundance measurements of $^4$He is much lower than the constraints obtained by fitting the Planck18 CMB data. Since these different $\Omega_{\rm g}$ values influence cosmic dynamics at vastly different epochs, it raises the possibility of a logarithmic running of the cosmic glitch with scale, analogous to how other dimensionless constants behave in renormalizable theories. In \ref{fig:running}, we illustrate the constraints on the glitch parameter~$\Omega_\mathrm{g}$ from BBN and CMB data, along with the potential linear extrapolations based on renormalization group flow. The BBN and CMB constraints suggest a possible scenario where the glitch almost vanishes and we recover near-exact de~Sitter symmetry on the scale of the observed cosmological constant (or de~Sitter radius) today. In stark contrast, at the Big Bang, where the curvature approaches the Planck scale, the glitch parameter becomes $\mathcal{O}(1)$, signaling a substantial violation of Lorentz symmetry potentially due to the quantum nature of gravity. Such a scenario may further hint at a genuine quantum gravity solution to the cosmological horizon problem, which is traditionally tackled under an inflationary paradigm \cite{afshordi2016critical}.

\begin{figure}[tbp]
    \centering
    \includegraphics[width=0.8\textwidth]{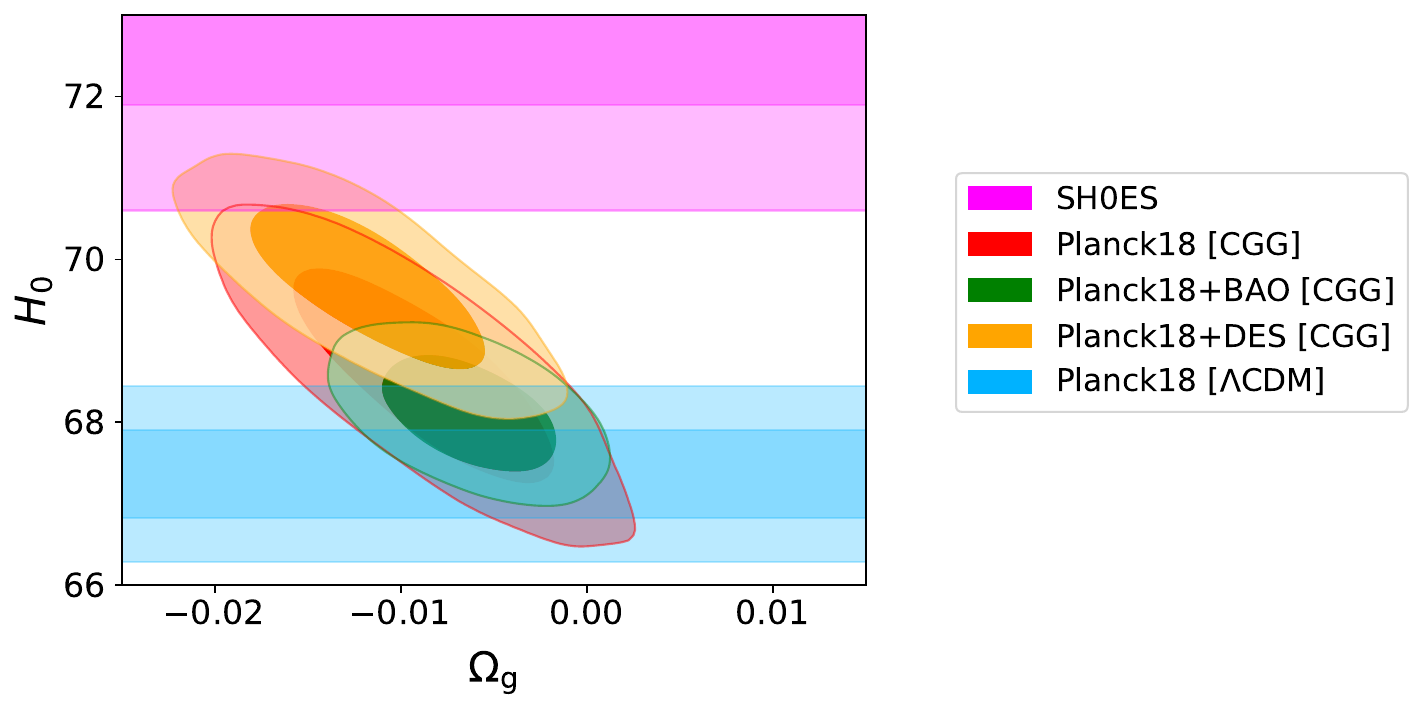}
    \caption[Caption for Hubble]{Constraints on the Hubble parameter~$H_0$ and glitch parameter~$\Omega_\mathrm{g}$ in CGG ($\Lambda$CDM+$\Omega_\mathrm{g}$) under different scenarios. The magenta band gives the distance-ladder measurement of $H_0$ from \textsc{SH0ES}~\cite{21SH0ES}, while the blue band shows the measurement of $H_0$ from Planck18 data under the $\Lambda$CDM model (i.e.\ setting $\Omega_\mathrm{g}=0$).}
    \label{fig:CGG-H0}
\end{figure}

\subsection{Cosmic tensions}

In addition to the discrepancy in the abundance of $^4$He, the increased precision of cosmological measurements has uncovered several other discrepancies among different cosmological probes~\cite{22CosmoIntertwined}, with the most notable being the so-called ``Hubble tension''. This tension refers to the difference between the measurements of the present-day cosmic expansion rate~$H_0$ based on the distance ladder in the late Universe and those obtained from CMB anisotropies analyzed through the standard cosmological model. Specifically, the discrepancy between the value inferred from the \textit{Planck} data~\cite{18Plancklikelihood,18Planckparameter,Tristram2023} under the $\Lambda$CDM model and the value from the SH0ES collaboration using Cepheid-calibrated Type Ia supernovae~\cite{2019SH0ES,21SH0ES,22SH0ES} has now surpassed the significance level of $4\,\sigma$. By exploiting the degeneracy between $\Omega_\mathrm{g}$ and $H_0$, where a negative $\Omega_\mathrm{g}$ allows for a lower matter density and a higher $H_0$, our CGG model can somewhat alleviate the Hubble tension. We determine $H_0$ to be $68.58\pm0.86~$km/s/Mpc for our CGG model under Planck18, which is higher than the $\Lambda$CDM value and has a larger uncertainty, thus mildly decreasing the Hubble tension from $4.1\,\sigma$ to $3.0\,\sigma$. 

Adding baryonic acoustic oscillation~(BAO) measurements~\cite{11sixdf_2011_bao,15sdss_dr7_mgs,17BAODR12} from the previous Stage-III galaxy surveys to the Planck18 data leads to $H_0=68.11\pm0.46$~km/s/Mpc and $\Omega_\mathrm{g}=-0.0063\pm0.0031$. The Hubble parameter measured in this case is closer to the $\Lambda$CDM result than the CGG fit under the Planck18 data only, despite the addition of the BAO data slightly strengthening the evidence for the glitch. \hl{Substituting the default Planck18 likelihoods with a newer analysis of the Planck Public Release 4 (PR4) data \cite{20Npipe,Tristram2023} and replacing the Stage-III BAO measurements with the newer BAO results coming from the recent DESI Y1 analysis \cite{24DESI_BAO_measure,24DESI_BAO_constraint},\footnote{Note that the DESI Y1 BAO measurements have comparable (if not slightly better) constraining power to the previous stage-III galaxy surveys, and they will significant exceed the statistical power of previous results in the upcoming data releases.} the glitch and Hubble parameters slightly tighten to $\Omega_{\rm g}=-0.0067\pm 0.0029$ and $H_0=68.49\pm0.41$~ km/s/Mpc, respectively. The combined CMB and BAO constraints are comparable across these two different combinations.} Including the $3\times 2$-point analysis (combined galaxy clustering and weak gravitational lensing measurements based on photometric surveys) from the Dark Energy Survey~(DES)~\cite{18DES_Y1} along with the Planck18 data yields $H_0=69.69\pm0.66$~km/s/Mpc and $\Omega_\mathrm{g}=-0.0118\pm0.0042$. This further raises the fitted $H_0$ values to within $2.4\,\sigma$ of the local SH0ES measurement. These constraints on the Hubble parameter are summarized in~\ref{fig:CGG-H0}.

\begin{figure}[tbp] 
    \centering
    \includegraphics[width=0.7\textwidth]{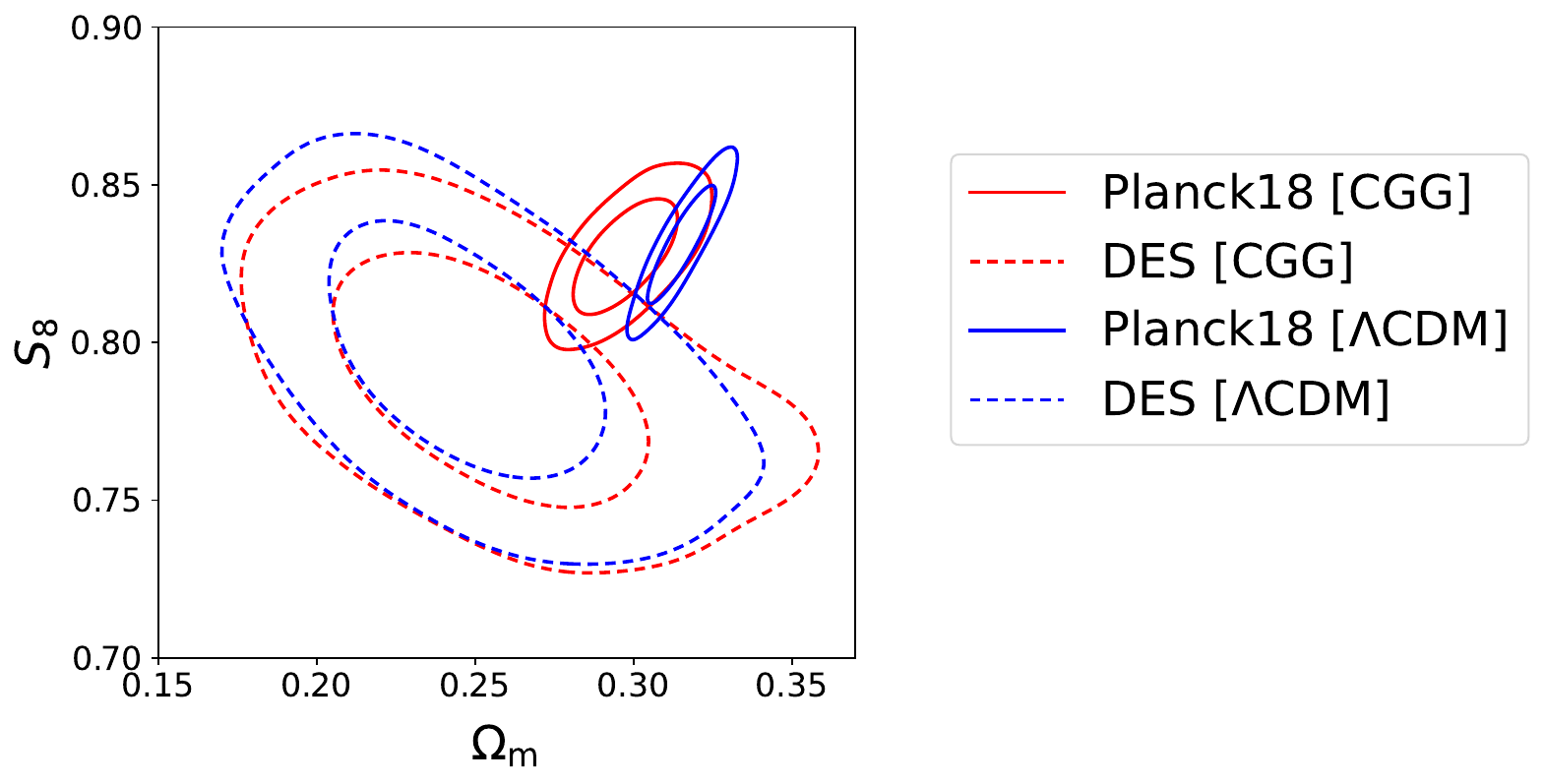}
    \caption{Constraints on $S_8$ and $\Omega_\mathrm{m}$ in the $\Lambda$CDM and CGG models, using Planck18 only and DES data only.}
    \label{fig:CGG-S8}
\end{figure}

Another tension related to the matter clustering in the Universe is found between galaxy clustering or weak lensing surveys and CMB measurements~\cite{22CosmoIntertwined,19Handley}. This clustering tension is typically characterized through the matter density parameter~$\Omega_\mathrm{m}$ and the matter power spectrum amplitude parameter~$\sigma_8$, or their combination~$S_8\equiv\sigma_8(\Omega_\mathrm{m}/0.3)^{0.5}$. We plot the constraints on $\Omega_\mathrm{m}$ and $S_8$ under Planck18 and DES data in~\ref{fig:CGG-S8}. Allowing $\Omega_\mathrm{g}$ to be negative in the CGG model lowers the constraints for $\Omega_\mathrm{m}$ under the Planck18 data. Consequently, the $S_8$--$\Omega_\mathrm{m}$ confidence regions of DES and Planck18 exhibit a greater area of overlap under the CGG model in comparison to $\Lambda$CDM, which is illustrated in \ref{fig:CGG-S8}. The clustering tension is therefore reduced by the CGG model when we view the tension in the 2D plane of $S_8$ and $\Omega_\mathrm{m}$ parameters. The clustering tension is similarly reduced in the $\sigma_8$--$\Omega_\mathrm{m}$ plane. 

\begin{table}[tbp]
\setlength{\tabcolsep}{10pt}
\begin{center}
\begin{tabular}{l c c}
\hline
\hline
\noalign{\vskip 1pt}
Likelihoods & $\Omega_{\rm g}$ Constraints  \\
\hline
\noalign{\vskip 2pt}
Planck18 [PR3] & $-0.0087\pm0.0046$ \\
Planck18+Pantheon SNe & $-0.0083 \pm 0.0041$ \\
Planck18+DES Y1 & $-0.0118\pm0.0042$\\
Planck18+Stage-III BAO & $-0.0063\pm0.0031$ \\
Planck PR4 & $-0.0054 \pm  0.0042$ \\
Planck PR4+DESI Y1 BAO & $-0.0067\pm 0.0029$ \\
All & $-0.0059\pm 0.0027$\\
\hline
\end{tabular}
\end{center}
\caption{\hl{Mean and $1\,\sigma$ uncertainties of the cosmic glitch parameter $\Omega_{\rm g}$ under different combinations of data and likelihoods. For the last row, the "All" case includes the combination Planck PR4, DESI Y1 BAO,  Pantheon+ SNe (supernovae), DES Y1, and SH0ES likelihoods.}}
\label{tab:Omegag_Constraints}
\end{table}


\section{Conclusions}

In summary, we have studied a \emph{cosmic glitch in gravity}, i.e.\ a model characterized by gravity behaving differently on super-horizon and sub-horizon scales, from both theoretical and observational perspectives. We find that the \textit{Planck} data and a combination of various other data sets prefer a negative cosmic glitch parameter $\Omega_\mathrm{g}$. The significance of the evidence for this glitch varies between $1.3\,\sigma$ and $2.8\,\sigma$, depending on the additional LSS data included in the Bayesian analysis. {The currently tightest constraint on the glitch is $\Omega_{\rm g}=-0.0059\pm 0.0027$, which is obtained from combining all of the CMB (Planck PR4 \cite{Tristram2023}), BAO (DESI Y1 \cite{24DESI_BAO_measure}), Supernovae (Pantheon+ \cite{21Scolnic_PantheonP}), galaxy clustering and weak lensing (DES Y1 \cite{18DES_Y1}) data, along with the distance ladder (SH0ES \cite{21SH0ES}) measurements.} These constraints indicate that the current cosmological data mildly favor superhorizon gravity to be slightly weaker than the case in GR. The glitch model alleviates both the Hubble and the clustering tensions when considering the \textit{Planck}~2018 data, while the $H_0$ measurement obtained under Planck18 combined with DES~Y1 data becomes more consistent with the SH0ES measurement. The preference for a negative glitch parameter and the reduction of cosmic tensions motivate further study of the CGG model in the future, when increasingly larger volumes of cosmological data become available from multiple new surveys.

To assess how upcoming cosmological measurements can improve the constraints on the glitch parameter~$\Omega_\mathrm{g}$, we follow Ref.~\cite{21WenT0} and use the combined Fisher-information forecasts from a cosmic-variance limited CMB experiment and a \textit{Euclid}-like BAO survey~\cite{Euclid2011}. We forecast the $1\,\sigma$ error of the parameter to be below~$10^{-3}$, which is roughly a 4-fold reduction in uncertainty over the current results. The precision of the upcoming Stage~IV CMB ~\cite{16CMBS4} and LSS surveys~\cite{Euclid2011,19LSST,20Eifler_Roman,14SPHEREx} will allow us to determine whether $\Omega_\mathrm{g}$ is genuinely negative (or just due to a statistical fluctuation or unknown systematics in the current cosmological data sets) and shed light on whether the CGG model resolves some of the cosmic tensions witnessed by different probes. At a more fundamental level, future constraints on the cosmic glitch will provide a new window into the nature of the symmetries that govern gravitational dynamics across the Cosmos, from its fiery Big Bang to its glacial heat death.

\acknowledgement{This research was supported by the Natural Sciences and Engineering Research Council of Canada. NA is further supported by the Perimeter Institute for Theoretical Physics. Research at Perimeter Institute is supported in part by the Government of Canada through the Department of Innovation, Science and Economic Development Canada and by the Province of Ontario through the Ministry of Colleges and Universities. 
LTH was supported by a Killam Postdoctoral Fellowship and a CITA National Fellowship. Computing resources were provided by the Digital Research Alliance of Canada/Calcul Canada (\url{alliancecan.ca}). Parts of this paper are based on observations obtained with \textit{Planck} (\url{www.esa.int/Planck}), an ESA science mission with instruments and contributions directly funded by ESA Member States, NASA and Canada, and results from Dark Energy Spectroscopic Instrument (DESI, \url{www.desi.lbl.gov}). This paper used the codes \texttt{CAMB} (\url{camb.readthedocs.io/en/latest/}) and \texttt{Cobaya} (\url{cobaya.readthedocs.io/en/latest/}).
}

\bibliographystyle{spphys.bst}
\bibliography{refs.bib}
\end{document}